%% file: higgs1.tex
\documentclass[11pt]{article}
\usepackage{epsfig,sint,macros}
\begin{document}
\input title.tex
\input sect1.tex

\input sect2.tex
\input sect3.tex

\input sect4.tex
\input concl.tex

\input higgs1_bbl.tex
\end{document}

%% file: title.tex
\begin{titlepage}
\begin{flushright}
  DESY 98-088
\end{flushright}

\vskip 1 cm
\begin{center}
  {\Large\bf String breaking in SU(2) gauge theory \\
             with scalar matter fields }
\end{center}
\vskip 1 cm
\vbox{
\centerline{
\epsfxsize=2.5 true cm
\epsfbox{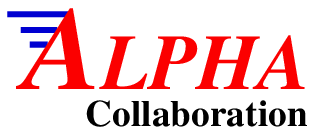}}
}
\vskip 1 cm
\begin{center}
{\large Francesco Knechtli and 
          Rainer Sommer  }
\vskip 1.0cm
DESY \\
Platanenallee 6, D-15738 Zeuthen
\end{center}
\vskip 2.5ex
{\bf Abstract}
\vskip 0.7ex
We investigate the static potential in the
confinement phase of the SU(2) Higgs model on the
lattice, where this model 
is expected to have properties similar to QCD.
We observe that Wilson loops are inadequate to
determine the potential at large distances, where the formation
of two color-neutral mesons is expected. Introducing smeared fields
and a suitable 
matrix correlation
function, we are able to overcome this difficulty. We
observe string breaking at a
distance
$r_{\rm b} \approx 1.8 r_0$, where the length scale $r_0$ has a value
$r_0 \approx 0.5\,\fm$ in QCD.
The method presented here
may lead the way
towards
a treatment of string breaking in QCD.
  \vfill

\begin{flushleft}
  DESY 98-088 \\
 July 1998
\end{flushleft} 

\eject

\vfill

\eject

\end{titlepage}

%% file: sect1.tex
\section{Introduction \label{s_intro}}

Since the seminal work by Creutz \cite{pot:creutz}, there have been a 
number of detailed studies of the static potential 
in non-Abelian gauge theories. At large distances, there is a 
linear confinement potential
between a  source anti-source pair
 in the fundamental 
representation of the gauge group.
This was clearly established by Monte Carlo calculations
of the lattice theories close to the continuum limit, both
for gauge groups 
$\SUtwo$ \cite{pot:UKQCDSU2,pot:r0} and $\SUthree$
\cite{pot:bali_first,pot:ukqcd65}.
When these gauge theories are coupled to matter fields in the fundamental
representation, one does expect that the potential flattens at 
large distances and asymptotically turns into a Yukawa form.
At such distances the potential
is better interpreted as the potential between static color-neutral
``mesons'', which are bound states of a static 
color source 
and the dynamical matter field. So far, this expectation could 
not be verified by Monte Carlo simulations. In particular, in recent 
attempts in QCD with two flavors of dynamical quarks this
string breaking effect was not visible 
\cite{pot:SESAM1,pot:guesken,pot:CPPACS,pot:UKQCDnf2}.
The same behavior of the potential is, of course, expected 
in the non-Abelian Higgs model in the confinement
phase. While the potential in the large distance range
could not be calculated in early simulations with gauge group $\SUtwo$,
they yielded
some qualitative evidence for screening of the potential
\cite{Higgs:Aachen1,Higgs:Aachen2}.
Similarly, the potential between static {\em adjoint} charges is
screened by the gauge fields themselves. Numerical evidence for this
has been found for gauge group $\SUtwo$
\cite{adjpot:su2michael}.

In this letter we consider the $\SUtwo$ Higgs model (in four dimensions)
as a first test case and demonstrate that string breaking exists. 
In order to roughly compare the physical situation in the 
Higgs model with the one in QCD, we  
choose a common reference energy scale. One immediately thinks of using the 
string  tension. However, due to 
the phenomenon of string breaking itself, 
the string tension does not have a precise meaning
(there is no range of $r$ where the potential is linear)
and exists only in an approximate sense. It is
better to fix the overall scale by $r_0$, defined as \cite{pot:r0}
\bes
 \left. r^2 F(r)\right|_{r=r_0} = 1.65\,, \label{e_defr0}
\ees
where $F(r)$ denotes the force derived from the static potential discussed
above.\footnote{To be precise: in the theory with matter fields, $r^2 F(r)$
is not monotonic and we expect that there are two solutions for $r_0$. 
The smaller one is to be selected.} The number 
on the r.h.s. of
\eq{e_defr0} has been chosen such that
$r_0$ has a value (estimated from 
phenomenological potential models describing
charm- and bottom-onia) of
\bes
 \rnod\approx 0.5\,\fm
\ees
in QCD. In the following, all dimensionful quantities 
are measured in units of $\rnod$,
which is not related to any phenomenological value 
in the $\SUtwo$ Higgs model.
In QCD with light quarks,
the distance $r_{\rm b}$ around which the potential should start flattening 
off, could be estimated in the quenched approximation
\cite{reviews:beauty}:
\bes
  r_{\rm b}\approx2.7\,\rnod \,\,\, \mbox{(in QCD)}\,.
\ees

%% file: sect2.tex
\section{The Higgs model on the lattice \label{s_higgs}}

Working on a four-dimensional hyper-cubic lattice  
with spacing $a$ we denote
by $\Phi(x)$ a complex Higgs field 
in the fundamental representation of the gauge group SU(2)
and $U(x,\mu) \in {\rm SU(2)}$ is the gauge field
link connecting $x+a\muh$ with $x$. 
We use the Greek symbols, $\mu,\;\nu$, to denote all directions 0,1,2,3
and Roman symbols such as $k$ to denote the space-like directions 1,2,3.
The Euclidean action for the SU(2) Higgs Model is then given as
\bes\label{eact1} 
  S & = & \sum_x a^4 \{ \sum_{\mu}
  (\nabla_{\mu}\Phi(x))^{\dagger} \nabla_{\mu}\Phi(x) -
  m_0^2\Phi^{\dagger}(x)\Phi(x) + \lambda_0[\Phi^{\dagger}(x)\Phi(x)]^2
  \} \nonumber \\
  & & +\sum_p \beta\{1-\frac{1}{2}\tr U_p\} \, , \\
  U_p(x) & = & U(x,\mu)U(x+a\muh,\nu)
  U^{\dagger}(x+a\nuh,\mu)U^{\dagger}(x,\nu) \, , \quad 
  \sum_p\,=\,\sum_x\sum_{0\le\mu<\nu\le 3} \, , \nonumber
\ees
where we used the lattice covariant derivative, 
$
  \nabla_{\mu}\Phi(x)  =  \frac{1}{a}\left[ 
  U(x,\mu)\Phi(x+a\muh) - \Phi(x) \right] 
$.
Following conventional notation \cite{Higgs:Montvay1},
the bare parameters chosen in the present investigation
are given by
\bes\label{parameter}
  \beta&=&2.2 \,,\nonumber \\ 
  \kappa & = & (1-2\lambda)/(8-a^2m_0^2) \,=\,0.274 \,,\\
  \lambda & = & \kappa^2\lambda_0 \,=\,0.5 \, . \nonumber
\ees
This point in parameter space is in the confinement phase,
fairly close to the phase transition, where the model
has properties similar to QCD \cite{Higgs:Montvay1,Higgs:Aachen1}. 
The lattice resolution is of roughly the same size as the one 
used in the QCD-studies 
quoted above:
from the potential calculation detailed in the following
section, we obtained
\bes\label{r0higgs}
  r_0 / a & = & 2.83 \pm 0.03 \, .
\ees
While one cannot expect a very precise result for the detailed form of the
potential with such a resolution, it is expected that qualitative 
features are correctly described. 

The string breaking distance $r_{\rm b}$ depends 
directly on the mass of the static mesons which in turn is  
sensitive to the bare 
mass, $m_0$, of the scalar fields. For our chosen value of $m_0$, 
the string breaking distance, $r_{\rm b}$, is somewhat smaller than the
estimate quoted above for quenched QCD.  Nevertheless, the 
physical situation is quite similar. We note in passing that the 
self-coupling of the Higgs field, $\lambda_0$, appears to have little 
influence on the physics of the Higgs model in the confinement 
phase \cite{Higgs:Montvay1}.

%% file: sect3.tex
\section{Calculation of the potential at all distances \label{s_pot}}

We now introduce a method, which -- as we will demonstrate in
the following section -- allows to compute 
the potential, $V(r)$, at all relevant distances in the 
theory with matter fields. 
Before explaining the details, we would like to mention
the basic point, which has first been noted by C. Michael
\cite{adjpot:su2michael}. Mathematically, the method is based
on the existence of the transfer matrix \cite{Luscher:TM} and 
the fact that it 
can be employed also when external static sources are present 
(see e.g. \cite{Borgs_Seiler}). 
In the path integral, a static source at position $\bf x$, together
with an anti-source at position ${\bf x}_r\,=\,{\bf x}+r\hat{k}$,
are represented by straight time-like Wilson lines fixed at these
space-positions. These Wilson lines have to be present in any (matrix) 
correlation
function from which one wants to compute the potential energy of
these charges. The space-like parts of the correlation functions,
which are again Wilson lines  when one considers standard Wilson loops, 
do not determine which states appear in the spectral representation of
the correlation functions.
They do, however, influence the weight with which different states
contribute. For these space-like parts,
we therefore use both Wilson lines which will
have large overlap with string-like states and Higgs fields
with a dominant overlap  with meson-type states. Combining them in a 
matrix correlation
function, the correct linear combination which gives the ground state
in the presence of charges can be found systematically
by the variational method described below. 

\begin{figure}[tb]
\hspace{0cm}
\vspace{-1.0cm}
\centerline{\epsfig{file=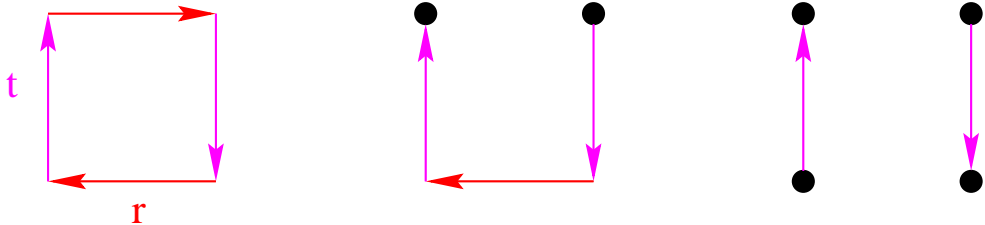,width=13cm}}
\vspace{-0.0cm}
\caption{The correlation functions used to determine the static
potential, from left to right: $C_{\rm WW}$, $C_{\rm WM}$ and 
$C_{\rm MM}$. The lines represent the Wilson lines, 
the filled circles the Higgs field.
\label{f_corr}}
\end{figure}

Let us now give precise definitions of
the correlation functions,
which are illustrated in \fig{f_corr}.
For small values of $r$ or in the pure
gauge theory, 
the static potential can be efficiently computed by means of Wilson loops,
\bes\label{wilsonloop}
  C_{\rm WW}(r,t) & = & \langle \tr[W(x,x_r)\, 
  W(x_r,x_r+t\hat{0})\,W^{\dagger}(x+t\hat{0},x_r+t\hat{0})\,
  W^{\dagger}(x,x+t\hat{0})] \rangle \nonumber \\ 
  & \simttoinfty & {\rm const.} \times \rme^{-V(r)t} \, ,
\ees
where $x_r\,=\,x+r\hat{k}$ and $W(x,y)$ denotes the product of 
gauge links along the straight line connecting $y$ with $x$. 
In addition, one may compute  
the mass, $\mu$, 
of a static meson from the correlation function
\bes\label{mesonmass}
  C_{\rm M}(t) =  \langle \Phi^{\dagger}(x)\,
  W(x,x+t\hat 0)\,\Phi(x+t\hat 0) 
  \rangle\; \simttoinfty \; {\rm const.} \times \rme^{-\mu t} \, .
\ees
Consequently, we expect that for distances significantly larger
than $\rb$,  where the relevant states correspond to 
weakly interacting 
mesons, the potential is close to the  
value $\lim_{r\to \infty}V(r) = 2 \mu$ and 
can be extracted from the correlation function
\bes\label{potbigr}
  C_{\rm MM}(r,t)  =  \langle \Phi^{\dagger}(x+t\hat 0)
  W^{\dagger}(x,x+t\hat 0)\Phi(x) \,\,\,
  \Phi^{\dagger}(x_r)W(x_r,x_r+t\hat 0)\Phi(x_r+t\hat 0)
  \rangle \, .
\ees
In order to investigate all (and in particular the intermediate)
distances, we introduce a (real)
symmetric matrix correlation function $C_{ij}(r,t)$, $i,j \in \{\rm W,M\}$
with
\bes\label{potintr}
 C_{\rm WM}(r,t)  =  \langle \Phi^{\dagger}(x+t\hat 0)\,
 W^{\dagger}(x,x+t\hat 0)\,W(x,x_r) 
 \, W(x_r,x_r+t\hat 0)\,\Phi(x_r+t\hat 0) \rangle \, .
\ees
For fixed $r$, $V(r)$ is extracted from the 
matrix correlation in the following way  \cite{phaseshifts:LW}. 
One first solves the generalized eigenvalue problem,
\bes\label{genev}
  C(t)v_{\alpha}(t,t_0) & = & 
  \lambda_{\alpha}(t,t_0)C(t_0)v_{\alpha}(t,t_0) \quad 
  \lambda_{\alpha} > \lambda_{\alpha+1} \, .
\ees
According to a general lemma proven in \cite{phaseshifts:LW}, 
the ground state energy $V(r)\equiv V_0(r)$ and the excited states
energies $V_1(r),\;...$ are then given by 
\bes\label{potentials}
  a V_{\alpha}(r) & = & \ln(\lambda_{\alpha}(t-a,t_0) 
  /\lambda_{\alpha}(t,t_0)) +
  \rmO\left(\rme^{-(V_N(r) - V_{\alpha}(r)) t}\right) \, .
\ees
Here, $N$ is the rank of the matrix $C$. So far we have $N=2$ 
and for this small value
of $N$ the method would require large values of $t$
for the correction terms in \eq{potentials} to be negligible. 

To improve on this,
we introduce smeared
space-like links \cite{smear:ape}, setting the smearing
strength $\epsilon$ of \cite{smear:ape} to the numerical
value $\epsilon\,=\,1/4$. 
For the Higgs field we employ a smearing operator,
$\mathcal{S}$, defined as
\bes\label{smearophiggs}
  \mathcal{S}\Phi(x) & = & \mathcal{P}\{\mathcal{P}\Phi(x) + 
  \mathcal{P}\sum_{|x-y|=\sqrt{2}a \atop x_0=y_0}
  \overline{U}(x,y)\Phi(y) + 
  \mathcal{P}\sum_{|x-y|=\sqrt{3}a \atop x_0=y_0}
  \overline{U}(x,y)\Phi(y)\} \, ,
\ees
where $\mathcal{P}\Phi = \Phi/\sqrt{\Phi^{\dagger}\Phi}$
 and  $\overline{U}(x,y)$ represents
the average over the shortest link connections between $y$ and $x$.  
\footnote{The particular form of $\mathcal{S}$ was found to be 
efficient in a study of \eq{mesonmass} with various different 
types of smearing operators. Details of this study will
be published separately.}

For 
different smearing levels $m=1,...,N_{\Phi}$ we evaluate
the correlation functions introduced above with $\Phi(x)$ replaced by 
$\Phi^{(m)}(x) = \mathcal{S}^m\Phi(x)$ and similarly with
$m=0,1,...,N_{\rm U}$ smearing iterations of the space-like gauge
fields, where $m=0$ corresponds to the unsmeared gauge links.
This defines a correlation matrix $C$ with rank $N=N_{\Phi}+N_{\rm U}+1$.

\begin{figure}[tb]
\hspace{0cm}
\vspace{-1.0cm}
\centerline{\psfig{file=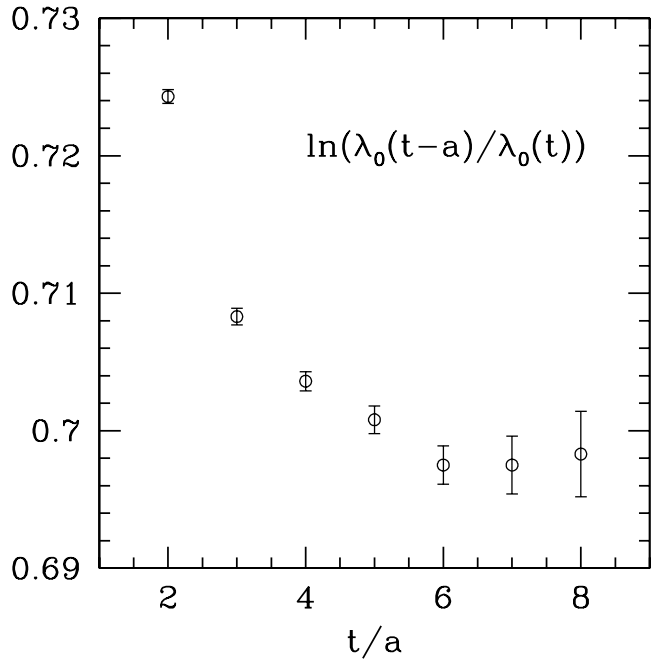,width=10cm}}
\vspace{-0.0cm}
\caption{Ground state energy of a static meson.
\label{f_meson}}
\end{figure}

%% file: sect4.tex
\section{String breaking \label{s_res}}
We now turn to discuss our numerical results obtained on
a $20^4$ lattice with periodic boundary conditions. The Higgs model can be 
simulated efficiently with a hybrid over--relaxation
algorithm. In particular, we implemented the
over--relaxation for the
Higgs field as described in \cite{Higgs:overrelax}. After a 
rough tuning of the mixture of the various parts of the updating 
we found that autocorrelations are no problem in 
our simulations. Errors were computed by a jacknife analysis.

We computed all
correlation functions up to 
  $r_{\rm max}=t_{\rm max}=8a \sim 3 \, \rnod$
on   2000 field configurations. We set $N_{\Phi}=4$ and $N_{\rm U}=2$.
The variance of the correlation functions was reduced in different ways.
We used translational invariance to average over the base
point denoted by $x$ in eqs.(\ref{wilsonloop}-\ref{potintr})
and cubic symmetry to average over the different orientations $\hat{k}$.
 Finally, statistical errors were further reduced by replacing 
-- wherever possible -- the 
time-like links by the 1-link integral
\cite{PPR}, which can be evaluated analytically
for the gauge group SU(2). 

\begin{figure}[tb]
\hspace{0cm}
\vspace{-1.0cm}
\centerline{\psfig{file=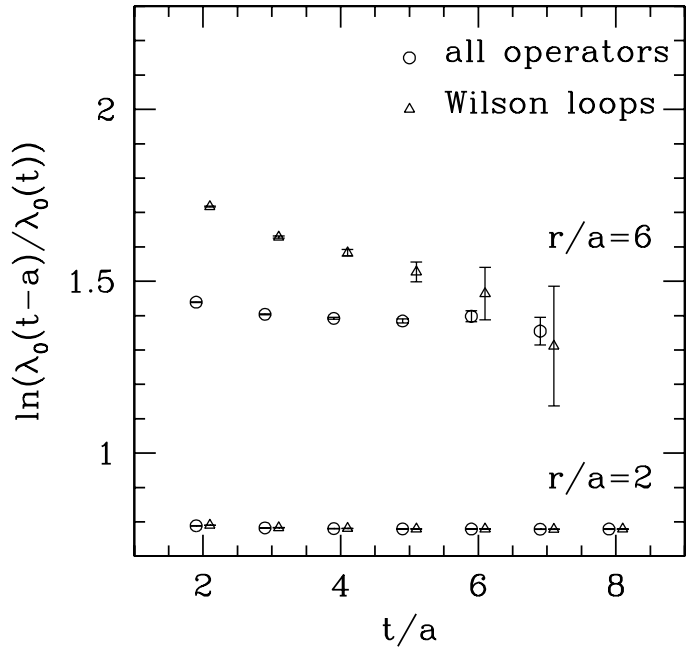,width=10cm}}
\vspace{-0.0cm}
\caption{Comparison of the static potential computed from
\eq{potentials} using the full correlation matrix and the sub-block
with the (smeared) Wilson loops. Two representative values of $r$
are shown.
\label{f_potential_t}}
\end{figure}

\begin{figure}[tb]
\hspace{0cm}
\vspace{-1.0cm}
\centerline{\psfig{file=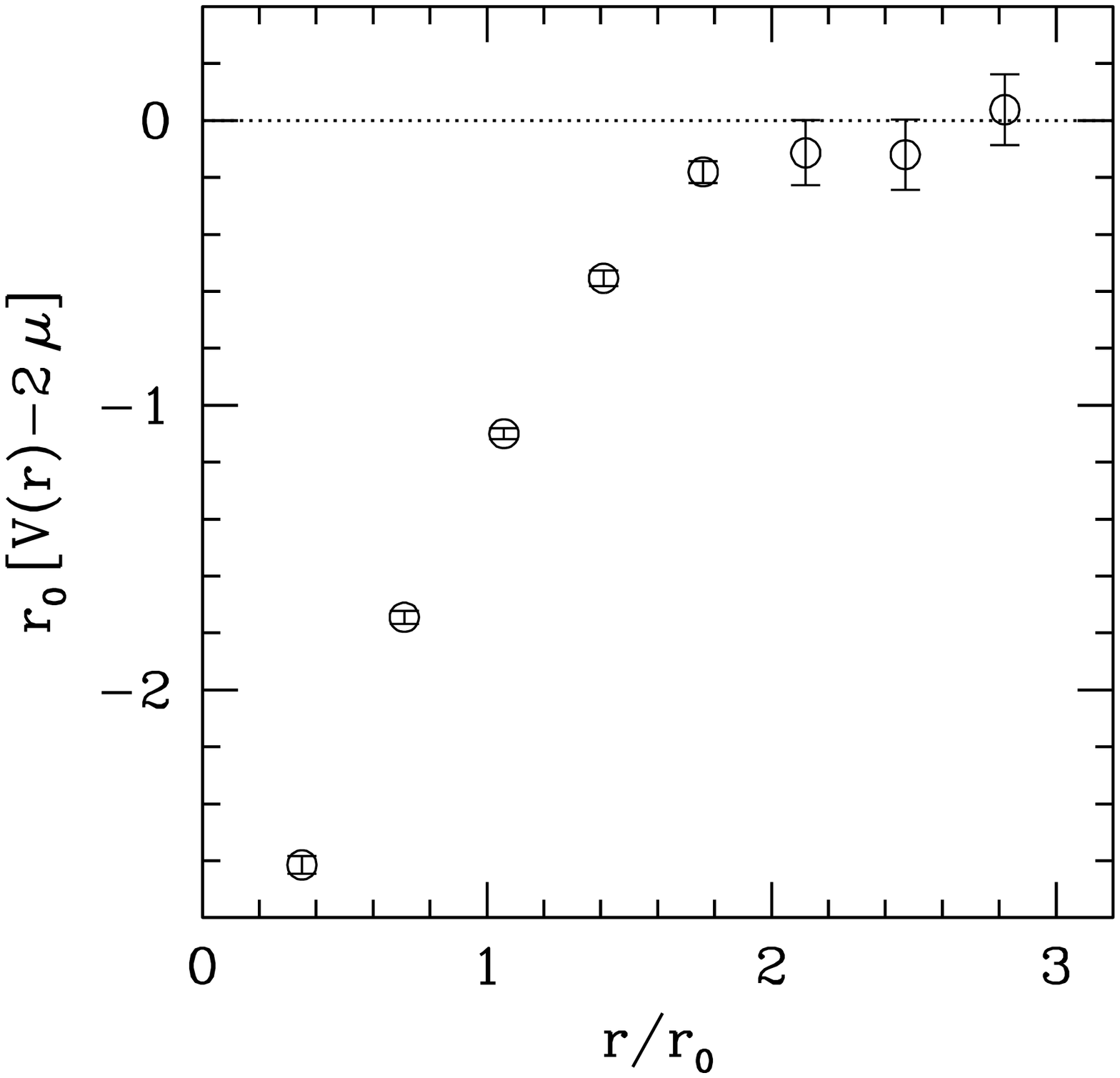,width=10cm}}
\vspace{-0.0cm}
\caption{The static potential in units of $\rnod$. The asymptotic
value $2\mu$ has been subtracted to obtain a quantity free of self 
energy contributions which diverge like $\frac{1}{a}$.
\label{f_invariant}}
\end{figure}

The first quantity one wants to know is the mass of the static mesons,
since this fixes the asymptotic value of the potential. It is best
computed from  the correlation $C_{\rm M}$ extended to a 
$N_{\Phi} \times N_{\Phi}$ correlation matrix by considering the
smeared $\Phi$ fields. As for all other energy levels discussed below,
the mass is computed from the eigenvalues of the generalized eigenvalue 
problem described above, setting $t_0=0$. In \fig{f_meson} the 
convergence for large $t$
is exemplified. One observes that the ground state energy can be 
extracted with confidence and with very good statistical precision,
$a\mu\,=\,0.698\pm 0.002$ (read off at $t=7a$ which agrees fully with 
$t=6a$). From the correlation
function without smearing a determination of $\mu$ was not possible.

At all distances,
the potential $V(r)$ was then computed 
using the full $7\times7$ correlation matrix. The convergence of
\eq{potentials} is shown in \fig{f_potential_t} (circles). 
In earlier calculations in QCD,
string breaking was searched for
in (smeared) Wilson loops. We studied whether one
can succeed in this way
by restricting the correlation matrix to the corresponding 
$(N_{\rm U}+1) \times (N_{\rm U}+1)$ sub-block. The resulting 
potential estimates (triangles in \fig{f_potential_t}) are
very good at short distances but have large correction
terms at long distances. Without a very careful analysis one might 
extract a potential which is too high at large distances, when
one uses Wilson loops alone. This {\em might} explain why string breaking 
has not been seen in QCD, yet. In contrast, from our full correlation matrix
 we can extract
safe estimates for $V(r)$ using \eq{potentials} for $t \approx 6a$. 


We then followed the steps described in \cite{pot:r0} to
determine $\rnod$ and computed the potential in units of $\rnod$. 
Considering in particular the combination $V(r)-2\mu$, one has
a quantity free of divergent self energy contributions. It is shown in
\fig{f_invariant}. The expected string breaking is clearly observed
for distances
$r>\rb\approx 1.8\,\rnod$. 
Around $r\approx\rb$, the potential changes rapidly from an almost
linear rise to an almost constant behavior. To resolve
this transition region using a smaller lattice
spacing is an interesting challenge to be addressed in the future.

{\bf Overlaps} of variationally determined wave functions $v_0$ 
are a certain measure for the efficiency of a basis of operators
used to construct the correlation functions. 
To give a precise definition of the overlap, we 
define the projected correlation function
\bes\label{projcorr}
  \Omega(t) \; = \; v_0^{\rm T}C(t)v_0 \; = \; 
  \sum_n \omega_n
  \rme^{-V_n(r) t} \, , \; \mbox{with normalization} \; \Omega(0)=1 \, .
\ees
Here, $n$ labels the states in the sector of the Hilbert
space with 2 static charges. The form given above
follows directly in the 
transfer matrix
formalism \cite{Luscher:TM}. 
The positive coefficients $\omega_n$ may be 
interpreted as the 
overlap of the true
eigenstates of the Hamiltonian with the approximate ground state 
characterized by $v_0$. ``The overlap '' is an abbreviation
commonly used to denote the ground state overlap, $\omega_0$.

\begin{figure}[tb]
\hspace{0cm}
\vspace{-1.0cm}
\centerline{\psfig{file=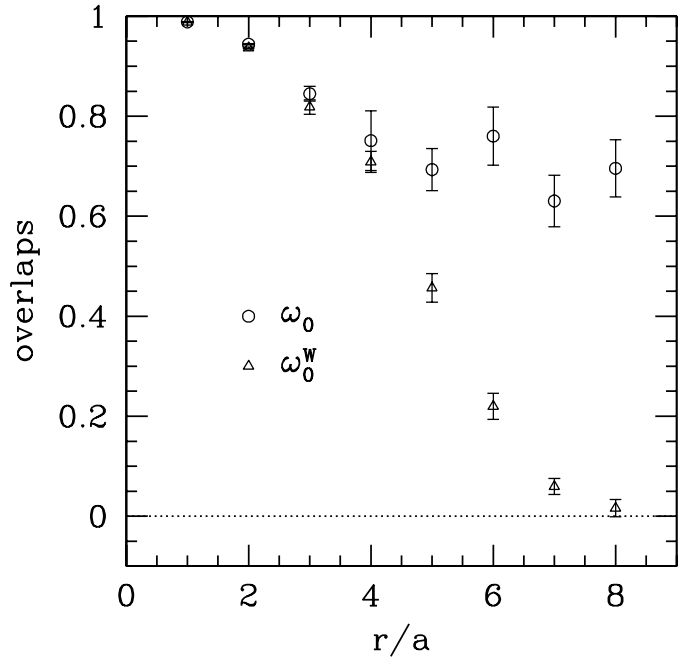,width=10cm}}
\vspace{-0.0cm}
\caption{Ground state overlaps, once
from the full correlation and once from the Wilson loops alone. 
\label{f_overlap}}
\end{figure}

We determine $\omega_0$ straightforwardly 
from the correlation function $\Omega(t)$ at large
$t$.  As shown by the circles in \fig{f_overlap}
our operator basis is big (and good) enough such that  
$\omega_0$ exceeds about 60\% for all distances.
It is now interesting to consider also
the overlap for the Wilson loops alone, i.e.
we again restrict the correlation matrix to the
$(N_{\rm U}+1) \times (N_{\rm U}+1)$ sub-block. 
Let us denote the corresponding projected correlation function by
$\Omega_{\rm W}$ and the overlap by $\omega_0^{\rm W}$. 
The computation of $\omega_0^{\rm W}$ is more difficult, because it
turns out to be very small at large $r$. Nevertheless, the 
expression 
\bes
  \omega_0^{\rm W} & \simttoinfty & \omega_0
  {\Omega_{\rm W}(t) \over \Omega(t)}
\ees
converges reasonably fast and $\omega_0^{\rm W}$ can be estimated
from the r.h.s. for $t \approx 6a$.
The results plotted in \fig{f_overlap} (triangles) show that
Wilson loops alone have an overlap which drops at intermediate distances and
are clearly inadequate to extract the ground state at large
$r$. Instead, the full matrix correlation function
has to be considered if one wants to compute $V(r)$ at all distances.

%% file: concl.tex
\section{Outlook \label{s_concl}}
We have introduced a method to compute 
the static potential at all relevant distances in gauge  
theories with scalar matter fields. We demonstrated that it 
can be applied successfully in the SU(2) Higgs model with parameters
chosen to resemble the situation in QCD.
There is little doubt that it can be applied
for different values of parameters in the Higgs model, at 
least as long as $\rb/a$ is not too 
large. It is then interesting to follow a line of constant physics
towards smaller lattice spacings in order to check for cutoff effects
and to be able to resolve the interesting transition region in the potential.
Furthermore, one might be interested in increasing the mass of the Higgs field 
in order to reach a situation with a larger value of $\rb/\rnod$
which presumably is even closer to the physics situation in QCD. 

From the matrix correlation function one can also
determine excited state energies.
We have done this successfully for the single meson states but
a precise determination of the excited potential at all distances
needs more statistics.
One expects that the transition region of the potential can be described
phenomenologically by a level crossing (as function of $r$) of the
``two meson state'' and the ``string state'' \cite{drummond:levelcross}. 
We are planning to investigate
this in more detail. So far, we can only say that for $r\approx\rb$
the two levels $V_1(r)$ and $V(r)$ are close.

Of course, it is of considerable interest to apply this method to QCD with 
dynamical fermions. It is difficult to predict how well this can be done.
Finding good smearing operators should not be a problem. However, the 
correlation functions will involve the quark fields and one cannot easily 
take advantage of translational invariance to reduce
the statistical errors as was done here. 
Thus, larger statistical uncertainties are expected. On the other 
hand, in QCD 
the quark fields are integrated out analytically, which usually
results in 
correlation functions with relatively small statistical errors. Also
new methods \cite{michael:alltoall} should be tried and
the final statistical errors can only be determined by explicit
computations. In any case, the only possible difficulty is expected
to be one of statistical accuracy. The proper correlation functions can be 
constructed along the lines of ref. \cite{adjpot:su2michael} and of
\sect{s_pot}.

The day of completion of this manuscript, a study of string breaking in   
the three-dimensional
SU(2) Higgs model appeared \cite{pot:higgs_3d}.
Both the method applied and the conclusions are very similar to
what we find in four dimensions.      

{\bf Acknowledgement.} We thank Jochen Heitger for helpful discussions.
Our simulations were performed on the SP2 of DESY at 
Zeuthen. We thank the staff of the DESY computer center for their support.